# An H- Surface Plasma Source for the ESS Storage Ring

**V. Dudnikov**

*Muons Inc.,*
*552 N. Batavia Ave. Batavia, IL 60510. USA*
*E-mail:* dvg43@yahoo.com

ABSTRACT: For H- charge exchange (stripping) injection to European Spallation Source (ESS) storage ring it is necessary to have H- ion source with a current of ~90 mA, pulses duration 3 ms, and repetition of 14 Hz (duty factor ~4%) extendable to repetition up to 28 Hz (df ~8%). Magnetron surface plasma H- source (SPS) with cesiation and active cathode and anode cooling is one of the possible solutions. Brookhaven National Laboratory (BNL) magnetron SPS can produce H- beam current of ~100 mA at ~2 kW discharge power and can operate up to a duty factor of ~1 % (average power ~14 W) without the active cooling. With active cathode and anode cooling it is possible to increase average discharge power up to 140 W(df 8%).For the delivery of the beam to Radio Frequency Quadrupole (RFQ) it is possible to use a short electrostatic Low Energy Beam Transport (LEBT) as in Spallation Neutron Source (SNS).

KEYWORDS: Surface plasma source; Negative ions; Magnetron; Discharge.

## Contents

**1. Introduction**
**2. Advanced Design Magnetron SPS**
**3. A Proposed ESS Injector**
**4. Conclusion**
**5. References**

## 1. Introduction

For H- charge exchange (stripping) injection [1,2] to European Spallation Source (ESS) storage ring it is necessary to have H- ion source with current ~90 mA, pulses 2.9 ms, and repetition of 14 Hz (duty factor ~4%) [3] extendable to repetition up to 28 Hz (df 8%). Magnetron surface plasma H- source with cesiation and active cathode and anode cooling is one of the possible solutions [4,5]. Brookhaven National Laboratory (BNL) magnetron Surface Plasma Source (SPS) can produce H- beam current of ~100 mA at ~2 kW (energy efficiency of H- generation up to 67 mA/kW, in Spallation Neutron Source (SNS) Radio Frequency (RF) SPS the energy efficiency ~1mA/kW) discharge power and can operate up to duty factor 0.7 % (average power ~14 W) without active cooling [6]. With active cathode and anode cooling it is possible to increase average discharge power up to 140 W (df 8%). In direct current cesiated H- ion source with Penning discharge [7] is used the discharge power of up to 1.5 kW. For beam delivery to Radio Frequency Quadrupole (RFQ) it is possible to use a short electrostatic Low Energy Beam Transport (LEBT) as in SNS [8].

## 2. Advanced design of Magnetron SPS

An advanced design of magnetron SPS with the spherical focusing of emitted negative ions and forced cathode and anode cooling is shown in Fig. 1. This new magnetron SPS is capable of DC operation with high average negative ion current generation.

A cross section of a new magnetron along the magnetic field is shown in Fig. 1a [9], and a cross section of a new magnetron perpendicular to the magnetic field is shown in Fig. 1b. Here a disc shape cathode (1) has diameter D= 18 mm and thickness H=12 mm. A surrounded anode (2) is separated from the cathode by an insulators (3). A vacuum gap between the cathode and the anode is d~1 mm. The cathode is cooled by liquid or gas flux flowing through the cooling tube (5) with OD~4 mm. The magnetron is compressed by ferromagnetic poles (4). A working gas is injected into the discharge chamber through the channel (10). Cesium is added to discharge through the second channel (11). The magnetic field, created by the magnet (13) and formed by magnetic poles (4) has a direction along the axis of the cooling tube (5).

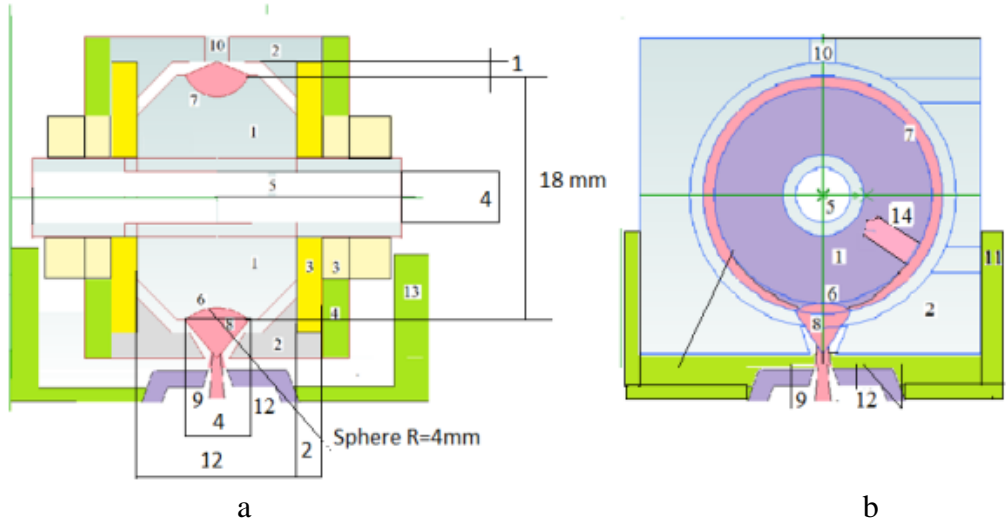

**Figure 1**. Cross sections of magnetron SPS with cathode cooling. (a-median cross section along to the magnetic field; b- cross section transverse to the magnetic field).1- cathode disc; 2-anode; 3-insulators; 4-magnetic poles; 5-cooling tube; 6-spherical dimple (negative ion emitter, R=4 mm); 7-cylindrical grove (discharge channel, r=3mm); 8-flux of focused negative ions; 9-negative ion beam extracted through emission aperture (2 mm diameter); 10- gas inlet; 11-cesium inlet; 12-extractor; 13- magnet; 14-hollow cathode.

The discharge in the crossed **ExB** fields is localized in the cylindrical grove (7) as in the semiplanotrons SPS [10,5]. The cylindrical grove focus emits negative ions to the anode surface and fast particles keep the anode surface clean by sputtering the flakes and deposits. A plasma drift in the discharge can be closed around the cathode perimeter or can be bracketed by a shallow cylindrical grove. Negative ions emitted from the spherical dimple (6) are used for beam formation, geometrically focused on the emission aperture made in the anode (2). These ions are extracted by an electric field applied between the anode (2) and the extractor (12). The emission aperture of ~2 mm diameter has a conical shape. The spherical dimple with a curvature radius R of ~4 mm has a working surface S of ~12 mm$^2$. For the emission current H- of 0.1 A it is necessary to have the emission current density on the cathode surface Je of ~ 1A/cm$^2$, which is acceptable for pulsed operation. The emission current density of H- in the amount of ~0.1 A/ cm$^2$ is necessary to achieve a 10 mA extraction acceptable for CW operation. Anode (2) is cooled by a flow of gas or liquid streaming through the cooling tube, attached to the front of the anode. Both cathode and anode used for H- beam production are to be made of Molybdenum. The surface of the spherical dimple must be mirror-smooth for efficient negative ion emission and be focused sharp into the emission aperture.

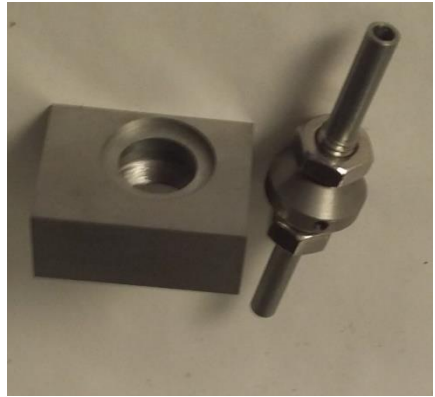

**Figure 2.** Photographs of cathode and anode of magnetro fn SPS with active cooling.

For heavy negative ion production, it is possible to use the cathode made of compound with a low work function as $LaB_6$ [11]. Two-stage extraction/acceleration is preferable for operation with high average beam current for collection of co-extracted electrons to the electrode with low potential. Gas valve [12] can be used for pulsed operation. A photograph of the cathode and anode of magnetron SPS with active cooling is shown in Fig.2.

## 3. A Proposed ESS Injector

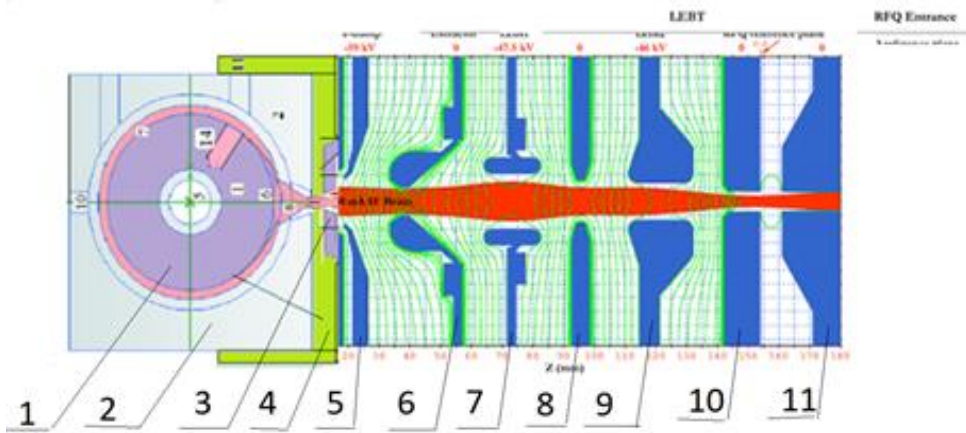

**Figure 3.** A schematic of proposed ESS injector. 1-cathode, 2-anode, 3-extractor, 4-magnetic pole, 5-electron damp, 6-grounded electrode, 7-lense 1, 8-grounded electrode, 9-lense 2, corrector, 10-RFQ wall, 11-RFQ van.

A schematic of the proposed ESS injector is shown in Fig.3. It consists of a surface plasma negative ion source with a magnetron configuration comprising of cathode 1 and anode 2 with emission aperture, extractor electrode 3 and magnetic pole 4. Extracted ion beam is accelerated to grounded electrode 6. Coextracted electrons are collected by the electron damp 4. The accelerated beam is focused by electrostatic Einzel lens 1 (7) and lens 2 (9) into RFQ wall aperture 10 and also by RFQ vanes.

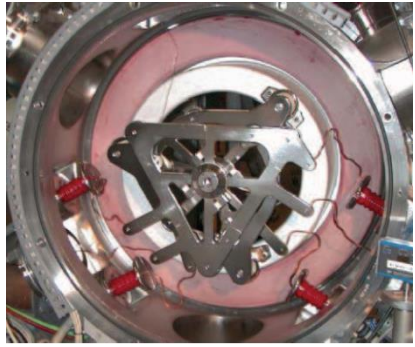

**Figure 4.** Construction of the LEBT for transporting H⁻ beam in RF SPS for SNS to the RFQ.

A design of electrostatic LEBT is shown in Fig.4 [13]. It operates well with H- beam current 60 mA at 65 kV with of up to 10%. Figure 5 shows the erosion of material on a BNL magnetron SPS that has successfully operated for 2 years: the cathode has a hole of 1.8 mm$^2$ close to the center of its spherical focusing dimple and the anode cover plate shows marks in the vicinity of the extraction hole spread across the area of 6.2 mm$^2$, which does not affect the magnetron operation. This damage is produced by the back-accelerated positive ions of $Cs^+$ and $H_2^+$. Estimation of sputtering of cathode and anode magnetron SPS was presented in [14].

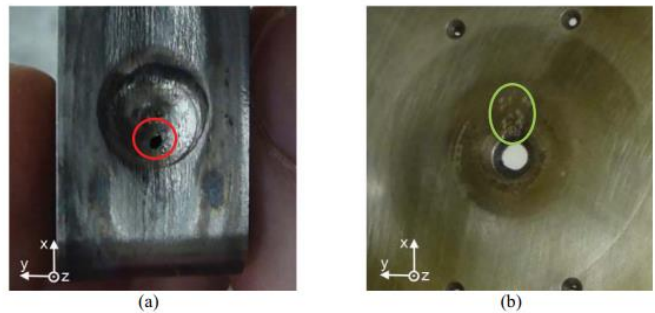

**Figure 5.** Deterioration of (a) the cathode and (b) the anode cover plate of BNL's magnetron. The location of the traces on the cathode and the anode cover plate is indicated by a circle and an ellipse (red and green),

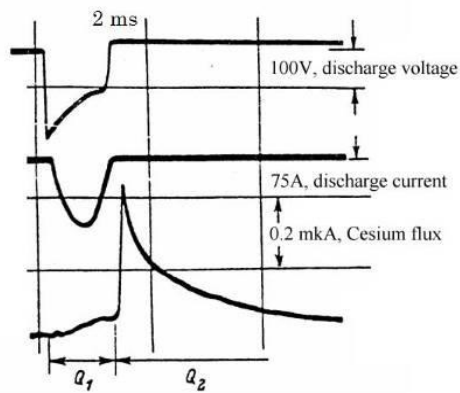

**Figure 6.** Characteristic oscillogram of Cesium ion current from mass spectrometer collector, illustrating the time variation of cesium atoms flux from a planotron SPS at high cathode temperature (~1000 K); also showing oscillograms of discharge current $I_d$ and discharge voltage $U_d$.

The estimation of Cesium density was incorrect because during discharge Cesium is strongly ionized and cannot escape the discharge chamber as shown in Fig. 6 from [5,[15]]. Fig. 6 shows a typical oscillogram of the Cesium ion current from the collector of the mass spectrometer, illustrating changes in the Cesium atoms flux from the source in time at a high (~1000 K) planotron cathode temperature, in conjunction with oscillograms of discharge current $I_p$ and discharge voltage $U_d$.[15]. One can see that cesium atoms leave the source mainly after the end of the discharge pulse. Cesium release during the pulse is small since Cesium is highly ionized and the extraction voltage blocks the escape of Cesium ions.

## 4. Conclusion

High current H- magnetron SPS with cesiation proposed for injector into ESS storage ring. The design of the magnetron SPS is discussed. Electrostatic LEBT is described. A lifetime of magnetron SPS is estimated as several Months.